\documentstyle[12pt]{article}
\newcommand{\pub}[4]{{\em #1 }{\bf #2}, #3 (#4)}
\newcommand{\prl}{Phys. Rev. Lett.}
\newcommand{\prb}{Phys. Rev. B}
\newcommand{\jpc}{J. Phys. C}

\begin{document}
\title{\large \bf Monte Carlo study of a vortex glass model}
\author{
{\normalsize J.~D.~Reger$^1$ and
A.~P.~Young$^2$}
\\
{\small \em $^1$
Institut f\"ur Physik, Universit\"at Mainz,
D-6500 Mainz, Germany}\\
{\small {\rm and} {\em AIX Competence Center, IBM Deutschland GmbH, D-8000 Munich,
Germany}} \\
{\small \em $^2$
Physics Department, University of California Santa Cruz,
Santa Cruz, CA 95064.}\\
}
\vfill
\date{}
\maketitle
\vfill
\begin{abstract}
We describe results of Monte Carlo simulations 
on a model that seems to have the necessary ingredients to
describe a disordered type-II superconductor in a magnetic field. We
compute the free energy cost to twist the direction of the phase of the
condensate and analyze the results by finite-size scaling. The results
show convincingly that the model has
different behavior as a function of dimension: in $d=4$
the model
clearly has a finite transition temperature; $T_c$,
while for $d=2$ only there is only a transition at $T=0$.
\end{abstract}
\vfill
\begin{center}
Submitted to {\em J. Phys. A}.
\end{center}
\newpage
\par
Since fluctuation effects play a much more important role
in high temperature superconductors than in conventional
superconducting materials, there has been a great deal of
effort \cite{ffh91} to understand the behavior
of type II superconductors in a magnetic field, including the effects
of disorder, when one goes beyond the mean field picture of BCS
or Ginzburg--Landau theories. One intriguing aspect which has emerged is
the possibility of a vortex glass phase \cite{mpaf89,ses84} in which the
off diagonal long range order of the pair condensate has a phase which
is random in space but frozen in time, much like the order parameter in
a spin glass \cite{by86}.
This can arise because the Abrikosov
flux lattice, which forms in pure samples, is
destroyed by disorder in less than 4 dimensions \cite{lo79} beyond a
certain length scale, $l_{dis}$.
The phase of the condensate does not then
form a regular periodic pattern on scales larger than $l_{dis}$, but,
according to the vortex glass hypothesis,
the system undergoes a transition into a spin glass--like
state in which the phase is frozen in time.
At the transition, the vortex glass correlation
length, $\xi$ diverges. A number of experiments \cite{kgk} have
found evidence for such a transition in the $I-V$ characteristics of 
Y-B-Cu-O samples.
Only if there is a vortex glass phase does the resistance really 
vanish \cite{ffh91} for
$H > H_{c_1}$. Otherwise, the resistance is, in principle,
finite because clusters of vortices on
scale $\xi$ can move by thermal activation over barriers, a process
known as ``flux creep'' \cite{ak}. 
These effects are observable \cite{tink88}
in high--$T_c$ compounds since they have much larger fluctuations than
conventional materials.

In this paper we descrive results of 
Monte Carlo simulations, analyzed by finite-size-scaling techniques,
on a model that seems to have the necessary ingredients to
describe the vortex glass state.  The results
show convincingly  the different behavior occurs in different
dimensions. In $d=4$ there is clearly a transition at finite transition
temperature $T_c$ with vortex glass order at lower temperature,
while in $d=2$ there is only a transition at $T=0$.
Analogous results for $d=3$ have been presented
before \cite{rtyf91}, and indicate behavior close to what is expected
at the lower critical dimension, though with probably a finite $T_c$.
Our results for $d=4$ have been briefly described in a
conference proceeding \cite{r92}. 
We feel
that it is useful to include them here as a contrast to the new results for
$d=2$ to emphasise the power of the finite-size-scaling technique in
elucidating whether or not a glass-like transition occurs.

The model that we study, known as the ``gauge glass'', has the following
Hamiltonian:
\begin{equation}
{\cal H} = - \sum_{<i,j>} \cos ( \phi_i - \phi_j - A_{ij}) \quad .
\label{ham}
\end{equation}
The phase, $\phi_i$, is defined on each site of a regular lattice, 
square for two dimensions, simple cubic for $d = 3$ and
simple hypercubic for $d = 4$, with $N = L^d$ sites. Periodic
boundary conditions are imposed.
The sum is over all nearest neighbor pairs on the
lattice. The effects of the magnetic field and disorder are
represented by the quenched vector potentials, $A_{ij}$, which we take to
be independent random variables with a uniform distribution
between 0 and $2\pi$.
This model seems to be the simplest
one with the correct ingredients of randomness, frustration and
order parameter symmetry.
It does, however, ignore screening, and therefore corresponds to an 
extreme type II limit in which $\kappa = \lambda / \xi \rightarrow \infty$,
where $\lambda$ is the penetration length.
Since $ \kappa \gg 1$ in the high $T_c$ superconductors, this limit is
not unreasonable. It is unclear, however, how much inclusion of screening via a
fluctuating gauge field would modify the behavior of Eq. (\ref{ham}).

If the $A_{ij}$ are restricted to the values 0 and $\pi$, the model
becomes the $XY$ spin glass, for which the lower critical dimension
is believed \cite{lcdxy} to be 4. However,
earlier work \cite{hs90,fty91,rtyf91},
has shown that the gauge glass is in a different universality class
from the $XY$ spin glass, presumably because it does not have the
the ``reflection'' symmetry, 
$\phi_{i} \rightarrow -\phi_{i} \ \forall \, i$ \cite{hs90}.

As discussed before \cite{rtyf91,fty91}, it is useful to consider the
change in free energy $\Delta F$ when one imposes a twist $\Theta$ along
one of the space directions, $x$ say. More precisely, the periodic
boundary conditions, $\phi_i = \phi_{i+L\hat{x}}$ are replaced by the
twisted boundary conditions, $\phi_i = \phi_{i+L\hat{x}} + \Theta$. By a
simple redefinition of the phases $\phi_i$ one can replace this situation by a
system with periodic boundary conditions and an extra contribution,
$\Theta / L$, to the vector potential on bonds in the $x$-direction.

By Monte Carlo methods one can
calculate derivatives of the free energy w.r.t. $\Theta$, so,
for a single sample, we define a current, $I$, and a stiffness, $Y$, by
\begin{equation}
I \equiv \frac{\partial F}{\partial \Theta}
 = \frac{1}{L} \sum_i \langle \sin \Delta_i \rangle_T \quad ,
\label{idef} 
\end{equation}
\begin{equation}
Y  \equiv \frac{\partial^2 F}{\partial \Theta^2} 
 = \frac{1}{L^2} \biggl \lbrace \sum_i \langle \cos \Delta_i  \rangle_T
- \ {1 \over T} \sum_{i,j} \Bigl [
\langle \sin \Delta_i \sin \Delta_j \rangle_T
- \ \langle \sin \Delta_i \rangle_T \langle  \sin \Delta_j 
\rangle_T \Bigr ] \biggr \rbrace \quad ,
\label{ydef} 
\end{equation}
where $\Delta_i = \phi_i - \phi_{i+\hat{x}} - A_{i, i+\hat{x}}$,
$F$ is the total free energy and
$i+\hat{x}$ refers to the nearest neighbor site in the
$x$-direction from $i$. Note that both $I$ and $Y$ are gauge invariant
so they are still useful even if one includes fluctuating gauge
fields.

Above $T_c$, $\Delta F$, and hence both $I$ and $Y$,
go to zero rapidly with increasing system
size because the system is insensitive to boundary conditions when $L$
is much greater than the vortex glass correlation length $\xi$.
If $T_c$ is finite, then, below $T_c, I$ and $Y$
vary with $L$ as $L^{\theta}$ where $\theta\ ( > 0)$, is an exponent
describing the low temperature phase. In other words, $I$ and $Y$
{\em increase} with
increasing $L$ below $T_c$,
the opposite of what happens above $T_c$. Precisely at
$T_c$, both $I$ and $Y$ are independent of size. Hence if $T_c$ is
finite, $I$ and $Y$ should come together at $T_c$ and splay out again at
lower temperatures. By contrast, if $T_c = 0 $, then, at $T = 0$,
$I$ and $Y$ vary as $L^\theta$ but with $\theta < 0$. Consequently, 
$I$ and $Y$ decrease with $L$ even at $T = 0$.

In a disordered system, it is necessary to perform an average over
different realizations of the disorder, which we indicate by
$[\cdots ]_{\rm av}$. For the gauge glass, the average values of $I$ and
$Y$ are both zero, i.e.
\begin{equation}
[ Y ]_{\rm av} = [ I ]_{\rm av} = 0 \quad ,
\end{equation}
because the configuration in which the vector potentials in the $x$
direction have been increased by $\Theta / L$
has the same weight in the configurational average as the
original choice of vector potentials. One is therefore interested in the
root mean square fluctutation between samples. This means that many
samples must be averaged over, typically several thousand. If $T_c$ is
finite, the finite
size scaling form for the r.m.s. current, $\Delta I$ is therefore
\begin{equation}
\label{finitetc}
\Delta I \equiv [ I^2 ]^{1/2}_{\rm av} = \tilde{I}(L^{1/\nu}(T - T_c)) \quad
(T_c > 0) \quad ,
\end{equation}
where $\nu$ is the correlation length exponent. We shall 
concentrate on the r.m.s. current in what follows, rather than the
stiffness, because sample to sample flucutations in the stiffness
have an asymmetric distribution with a long tail, which makes it
difficult to get good statistics \cite{rtyf91}.
If $T_c = 0$ then $\Delta I$ decreases with size even at $T=0$, i.e. $\Delta I
\sim L^\theta$ where $\theta$ is negative and related to the exponent
$\nu$ giving the divergence of the correlation length as $T \to 0$ by
$-\theta = 1 / \nu$. The finite size scaling form is then
\begin{equation}
\label{zerotc}
L^{1/\nu} \Delta I =  \tilde{I}(L^{1/\nu}T) \quad
(T_c = 0) \quad .
\end{equation}

Tests to ensure equilibration
were carried out as described elsewhere \cite{by88}. 

We first show results for $d = 4$, \cite{r92}.
Fig. 1 shows clearly that the data for the r.m.s. current for
different sizes come together at $T = T_c \simeq 0.95$ and then
splay out again on the low temperature side, just as expected at a
finite temperature transition. This
provides unambigous evidence that there is a vortex glass transition
in four dimensions (and presumably also in higher dimensions) with
vortex-glass order on the low-temperature side. A
scaling plot corresponding to Eq. (\ref{finitetc})
is shown in Fig. 2. From the fit we estimate
\begin{equation}
T_c = 0.96 \pm 0.01, \quad \nu = 0.7 \pm 0.15 \quad (d = 4) \quad .
\end{equation}

Next we discuss the case of $d = 2$. The results for $\Delta I$ are
shown in Fig. 3. Notice that they are quite different from Fig. 1, since,
even at the lowest temperature, $\Delta I$ decreases with increasing
size. This is precisely what is expected at a zero temperature
transition, and the scaling plot in Fig. 4 corresponding to Eq. (\ref{zerotc}) 
works very well. From the fit we estimate
\begin{equation}
T_c = 0 , \quad \nu = 2.2 \pm 0.2 \quad (d = 2) \quad .
\end{equation}
The value for $\nu$ agrees with earlier work \cite{fty91}, in which a different
finite-size-scaling technique was used.
Recent experiments \cite{dekker}
on very thin (16\AA) films of YBCO have provided striking confirmation
that
$T_c=0$ for the vortex glass in two dimensions. Furthermore, it is found
that non-linear current-voltage characteristics set in when
the current density exceeds a value, $J_{nl}$ which, according to
scaling theory \cite{fty91} varies with
temperature as $T^{1+\nu}$. According to conventional
flux creep theory, there is no divergent length scale as $T \to 0$,
which corresponds to setting $\nu=0$. The experiments \cite{dekker} find
$1 + \nu = 3.0 \pm 0.3$
in excellent agreement with the results
presented here and in Ref. \cite{fty91}, but in clear disagreement with
the flux creep theory. We emphasize, then, that the vortex glass
picture leads to measurable consequences even when $T_c = 0$.


To conclude, we have shown that a Monte Carlo calculation of the change in free
energy due to a twist in the boundary conditions combined with finite
size scaling is a very powerful tool for systems with $XY$-like
symmetry. We have shown that it clearly distiguishes between the finite
temperature transition in the $d=4$ gauge glass and the zero temperature
transition of the two-dimensional model.

{\bf Acknowledgments:}
The work of APY was supported in part by the NSF grant No.\ DMR 91-11576.
The work of JDR was supported in part by the
the Deutsche Forschungsgesellschaft through Sonderforschungsbereich 262/D1.
Parts of the computations were performed on the Cray--YMP 8/832 at the
German Supercomputer Center (HLRZ), J\"ulich.
It is a pleasure to acknowledge M.~P.~A.~Fisher for informative discussions.

\newpage
\section*{Figure Captions}
\begin{description}
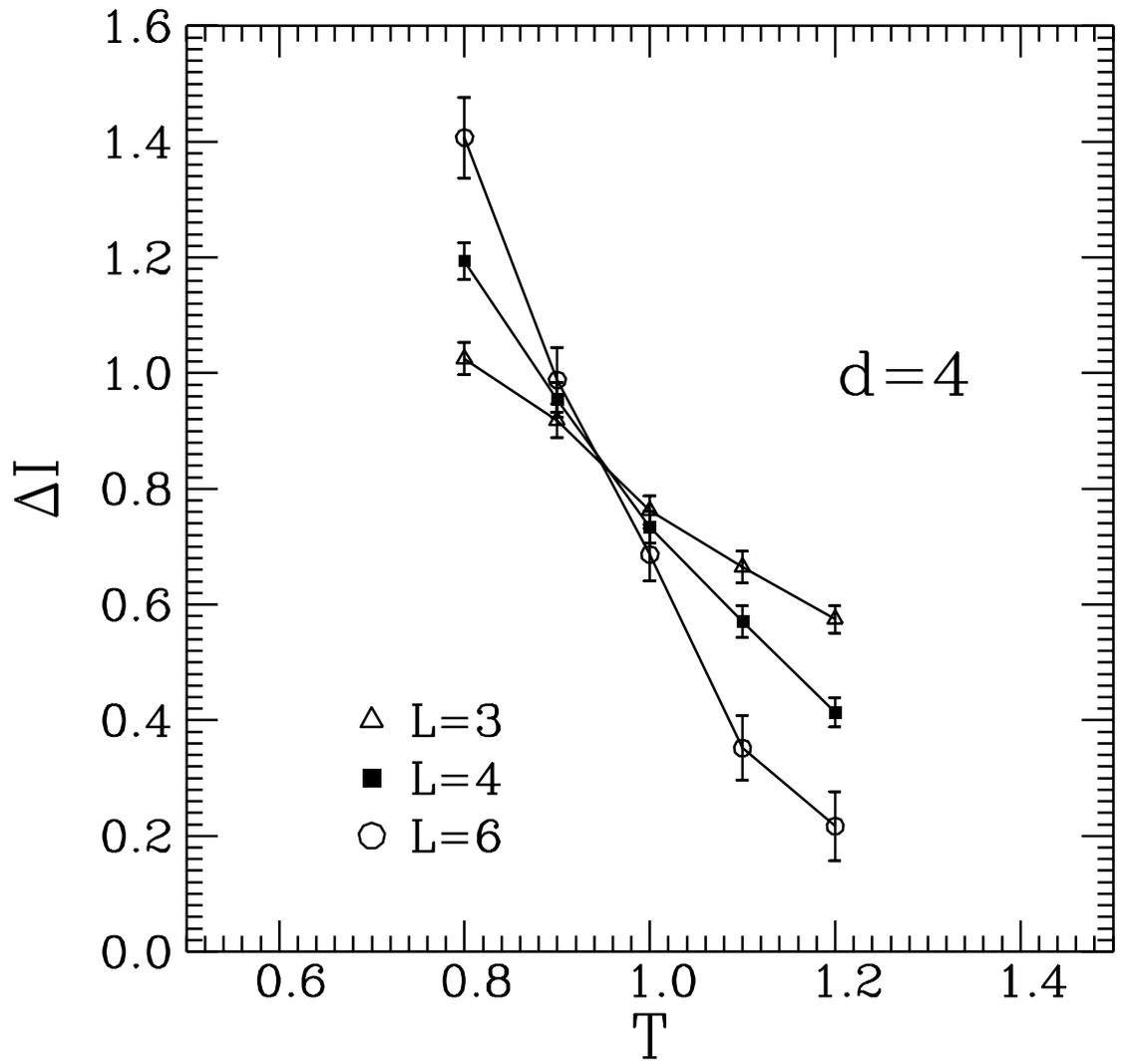
\item{\bf Figure 1:}
The r.m.s.\ current, $\Delta I = [ I^2]_{av}^{1/2}$, for $d = 4$
determined by Monte Carlo simulations for different
sizes and temperatures. The curves for different sizes
are expected to come together at
$T_c$ and, if there is order in the low temperature state,
to splay out again at lower temperatures. The data does indeed follow
this behavior.
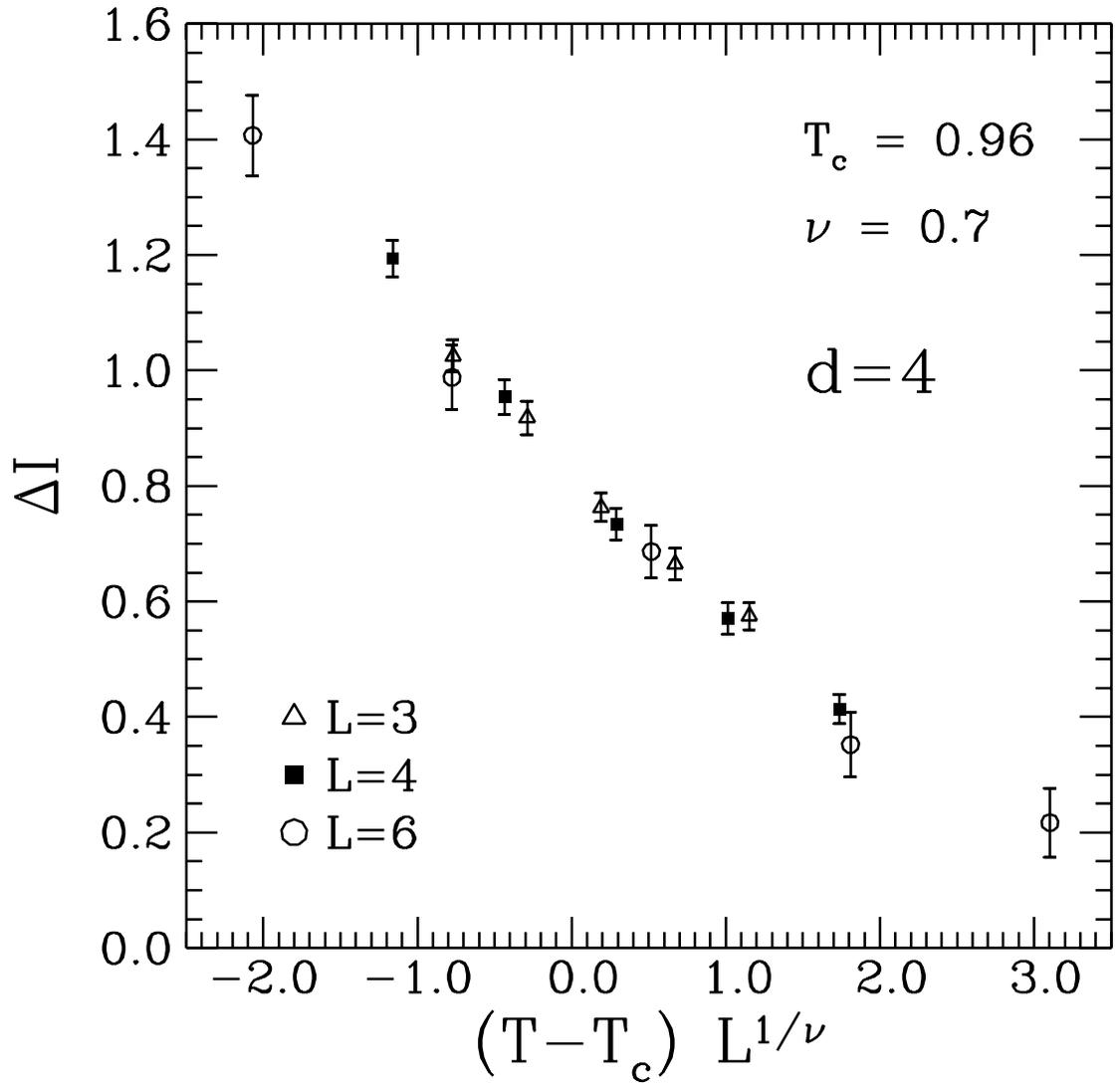
\item{\bf Figure 2:}
The same data as in Figure 1 but in a finite size scaling plot,
with $T_c = 0.96$ and $\nu = 0.7$.
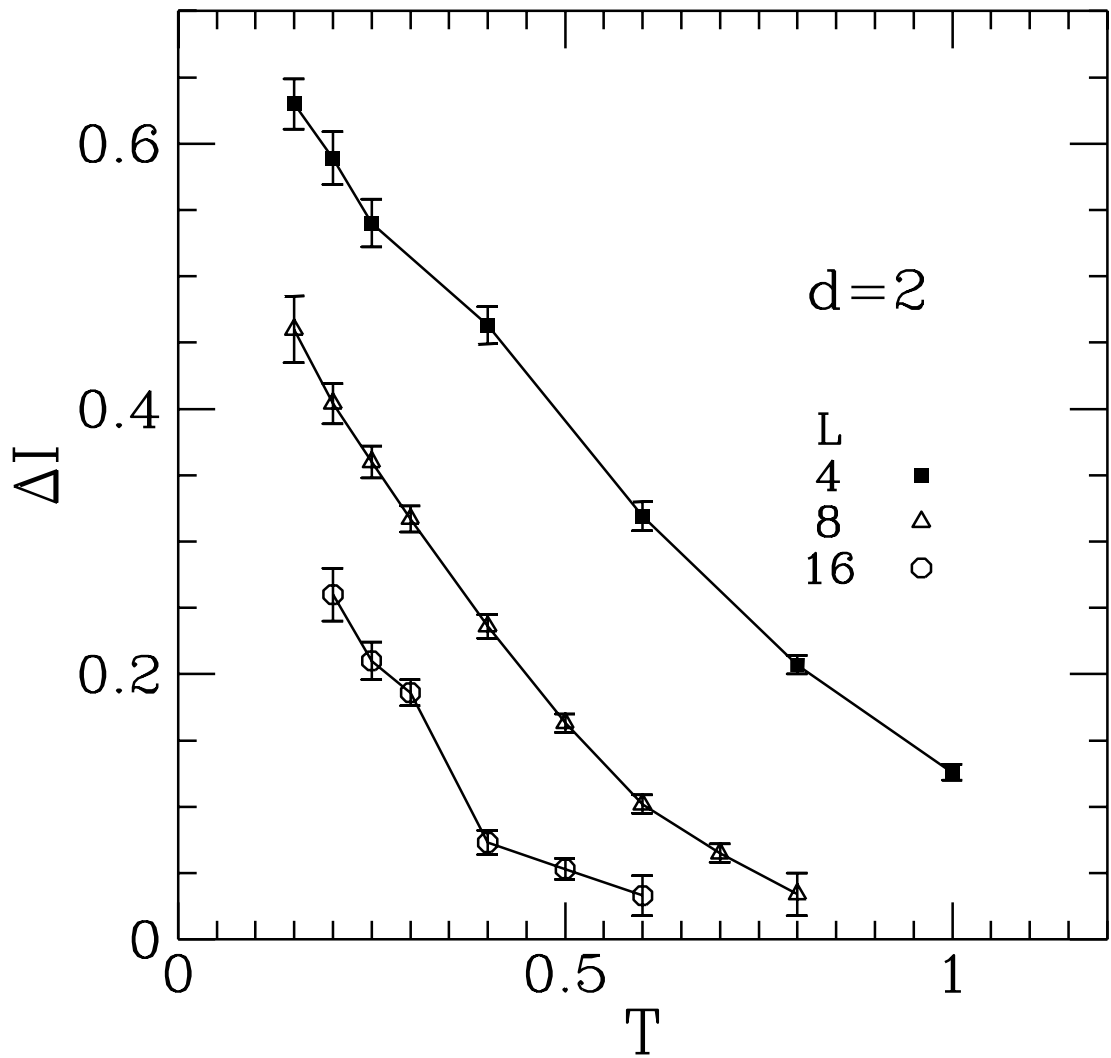
\item{\bf Figure 3:}
The r.m.s.\ current, $\Delta I = [ I^2]_{av}^{1/2}$, for $d = 2$
determined by Monte Carlo simulations for different
sizes and temperatures. The curves for different sizes
do not come together, even at the lowest temperature. This behavior
indicates a transition at $T = 0$.
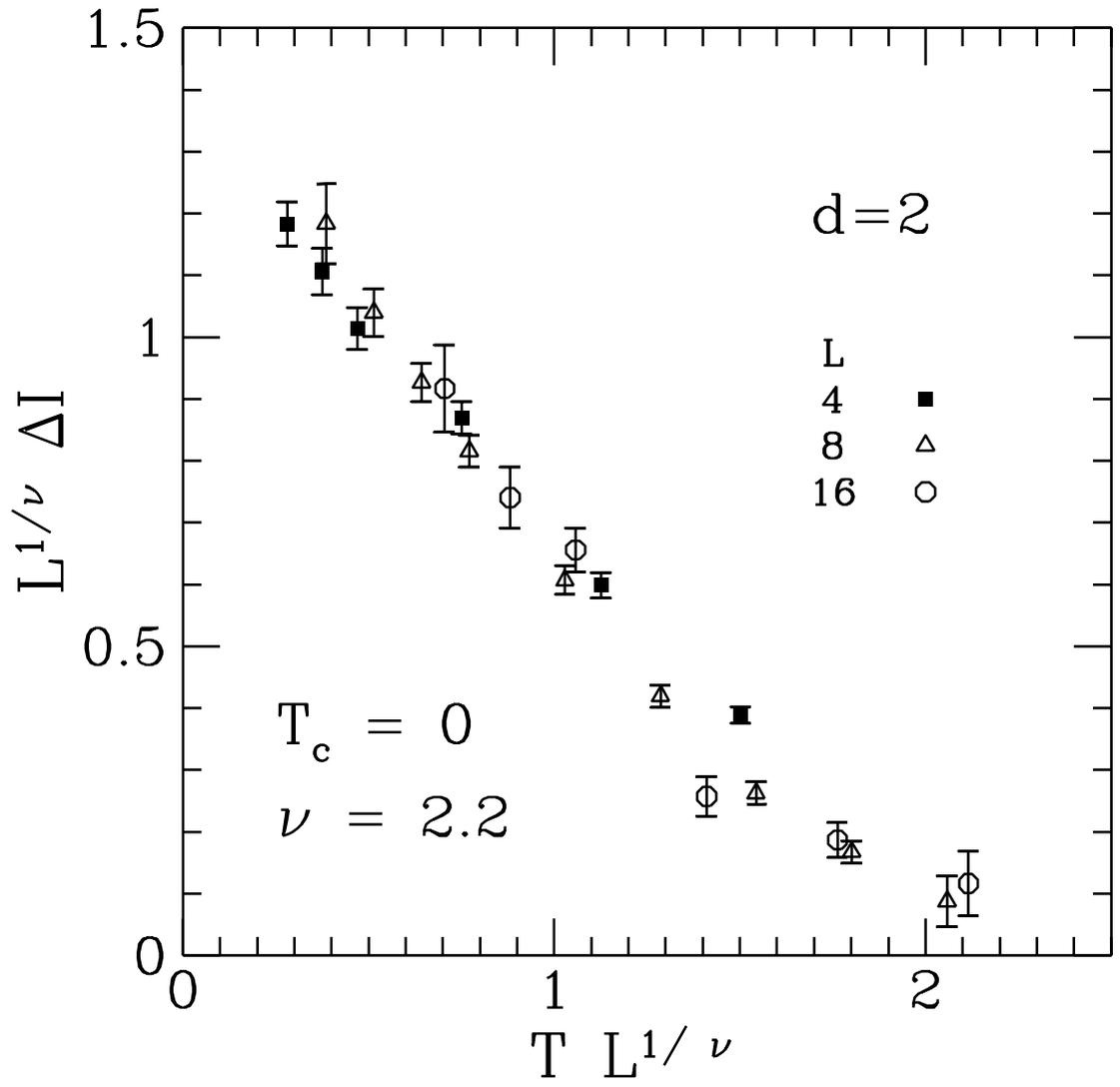
\item{\bf Figure 4:}
The same data as in Figure 4 but in a finite size scaling plot,
with $T_c = 0$ and $\nu = 2.2$.
\end{description}

\end{document}